# Reducing two-level system dissipations in 3D superconducting Niobium resonators by atomic layer deposition and high temperature heat treatment

Y. Kalboussi[1,a], B. Delatte[1], S. Bira[2,8], K. Dembele[3,3b], X. Li[4], F. Miserque[5], N. Brun[4], M. Walls[4], J.L. Maurice[3b], D. Dragoe[6], J. Leroy[7], D. Longuevergne[8], A. Gentils[8], S. Jublot-Leclerc[8], G. Jullien[1], F. Eozenou[1], M. Baudrier[1], L. Maurice[1], T. Proslier[1,a]

[1] *Université Paris-Saclay, CEA, Département des Accélérateurs, de la Cryogénie et du Magnétisme, 91191, Gif-sur-Yvette, France.*

[2] *Thomas Jefferson National Accelerator Facility, Newport News, 23606 Virginia, USA.*

[3] *Centre interdisciplinaire de microscopie électronique, Ecole Polytechnique, 91120 Palaiseau, France.*

[3b] *Laboratoire de physique des interfaces et couches minces, École Polytechnique, CNRS UMR7647, 91120 Palaiseau, France.*

[4] *Université Paris-Saclay, CNRS, Laboratoire de Physique des Solides, 91405 Orsay, France.*

[5] *Université Paris-Saclay, Service de Recherche sur la Corrosion et le Comportement des Matériaux, 91191 Gif sur Yvette, France.*

[6] *Université Paris-Saclay , Institut de chimie moléculaire et des matériaux d'Orsay, 91400 Orsay, France.*

[7] *Université Paris-Saclay, CEA, CNRS, NIMBE, LICSEN, 91191 Gif-sur-Yvette, France.*

[8] *Université Paris-Saclay, CNRS/IN2P3, IJCLab, 91405 Orsay, France.*

a) Authors to whom correspondence should be addressed: Yasmine.kalboussi@cea.fr and Thomas.proslier@cea.fr

## ABSTRACT

Superconducting qubits have arisen as a leading technology platform for quantum computing which is on the verge of revolutionizing the world's calculation capacities. Nonetheless, the fabrication of computationally reliable qubit circuits requires increasing the quantum coherence lifetimes, which are predominantly limited by the dissipations of two-level system (TLS) defects present in the thin superconducting film and the adjacent dielectric regions. In this paper, we demonstrate the reduction of two-level system losses in three-dimensional superconducting radio frequency (SRF) niobium resonators by atomic layer deposition (ALD) of a 10 nm aluminum oxide $Al_2O_3$ thin films followed by a high vacuum (HV) heat treatment at 650 °C for few hours.  By probing the effect of several heat treatments on $Al_2O_3$ – coated niobium samples by X-ray photoelectron spectroscopy (XPS) plus scanning and conventional high resolution transmission electron microscopy (STEM/HRTEM) coupled with electron energy loss spectroscopy (EELS) and  (EDX) , we witness a dissolution of niobium native oxides and the modification of the $Al_2O_3$-Nb interface, which correlates with the enhancement of the quality factor at low fields of two 1.3 GHz niobium cavities coated with 10 nm of $Al_2O_3$.

**THE MANUSCRIPT**

Superconducting radio frequency (SRF) resonators, historically used to accelerate particles and routinely achieving very high quality factors $Q > 10^{10}$ -$10^{11}$, are finding a new use in the quantum regime (at temperatures below 1 Kelvin) whether to be integrated into 3D quantum processing units [1,2] or to be used as quantum sensors to search for dark matter and gravitational waves [3]. The motivation behind these new applications is that SRF cavities benefit from a 1000-fold higher coherence lifetime than other 2D qubit architectures and can offer sensitivities orders of magnitude higher, which would greatly enhance many fundamental physics experiments [4]. Nonetheless, SRF cavities still suffer from dielectric losses arising from two-level systems (two-state defects) dissipations occurring in the native oxide layers that form once the superconductor is exposed to air. Regardless of their chemical composition, amorphous solids exhibit universal behavior at low temperature [5] caused by the presence of two-level system (TLS) defects within the material. Because of their low energy, such defects are saturated at high temperature. However, once the material is cooled to few kelvins, these additional degrees of freedom become available and are a major source of noise and decoherence in superconducting quantum devices [5]. Niobium, which is a commonly used material for these resonators, is known to grow an amorphous oxide layer once exposed to air. Recent work on three-dimensional 3D niobium (Nb) resonators in the low field regimes has shown that the niobium native oxide is a major source of TLS losses [4, 6] responsible for the degradation in the quality factors from $6.10^{10}$ to $2.10^{10}$ in 1.3 GHz cavities. The mitigation of TLS dissipations in 3D superconducting resonators is a fresh subject of research. So far, the only reported improvement in the quality factor of an SRF cavity in the quantum regime has been achieved by applying a high vacuum (HV) annealing step at 340 °C for 5 hours and keeping the cavity under vacuum to prevent its re-oxidation [7]. While this annealing step is effective, the necessity to sustain a vacuum environment makes it unpractical for quantum computing applications or quantum sensing. Another approach has been investigated by Bal *et al* [8] in which a metallic niobium film in a qubit structure has been encapsulated with an in-situ deposited tantalum film. This approach improved the coherence times by a factor of 2 to 5 but requires an oxide-free niobium surface to begin with.

In this paper, we investigate an approach combining an atomic layer deposition (ALD) coating on air-exposed niobium surfaces followed by a subsequent thermal treatment in HV to dissolve the initially present niobium native oxides. Such an approach was initially proposed by Proslier *et al* [9,10] in order to get rid of magnetic impurities present in niobium sub-oxides believed to limit the performances of accelerating cavities; the inner surface of the Nb cavity was coated with a thin protective layer of alumina followed by a subsequent annealing at 450 °C for 24 hours in HV. This resulted in the reduction of $Nb_2O_5$ into niobium sub-oxides by oxygen diffusion into the bulk Nb while the $Al_2O_3$ layer protected the metal layer from further oxidation. The corresponding RF tests showed an increase in Q due to the improvement of the superconducting surface properties but no investigation was made at that time of the low-field regime.

In this study, we opt for higher annealing temperatures, up to 650 °C, in order to promote further reduction of the initial native oxide and reduce the presence of sub-oxides. To this end, XPS measurements and TEM analysis were performed on air-exposed cavity-grade niobium coupons coated with an $Al_2O_3$ layer deposited by ALD. ALD is a self-limiting, sequential surface chemistry that has the unique ability to achieve uniform atomic-scale thickness control on complex-shaped substrates [11]. Pieces of polycrystalline Nb were cut from larger sheets used to fabricate SRF cavities. These pieces were electro-polished (EP) in a manner similar to that done on SRF cavities [12], cleaned in an ultrasonic bath and dried in air. They were later introduced into the ALD reactor and coated with 10 nm of $Al_2O_3$ at a temperature of 250 °C under a flow of ultra-pure nitrogen gas using a standard ALD recipe of 100 cycles of alternating Trimethylaluminium (TMA) and water $H_2O$ [13,14]. Both precursors were pulsed for 1 s and purged for 15 s. These deposition parameters resulted in a growth rate of 0.12 nm per cycle on niobium, in agreement with literature values [14]. More details on the ALD system used in this work can be found in [15]. The coated samples were then baked in HV (pressure $<10^{-6}$ mbar) at temperature of 650°C for several hours.

In order to assess this approach for the low-field performances of 3D Nb resonators, a 1.3 GHz niobium cavity was coated with a 10 nm $Al_2O_3$ film using the same deposition parameters used on the samples, and post-annealed in HV at 650°C for 4 hours. As a second test, we electropolished the niobium cavity to remove the $Al_2O_3$ layer and reset the niobium surface. We then tested the cavity to have a new baseline and performed the same $Al_2O_3$ coating; however, this time, we annealed the cavity at 650 °C for 10 hours. Prior to each RF test, the cavity was subject to a high pressure rinsing (HPR) with ultra-pure water [16]. The RF tests were conducted in the synergium vertical-testing facility at CEA. The low-field region quality factor has been measured following the cavity ring down procedure after shutting off RF power [17]. These tests are shown in Fig.1.

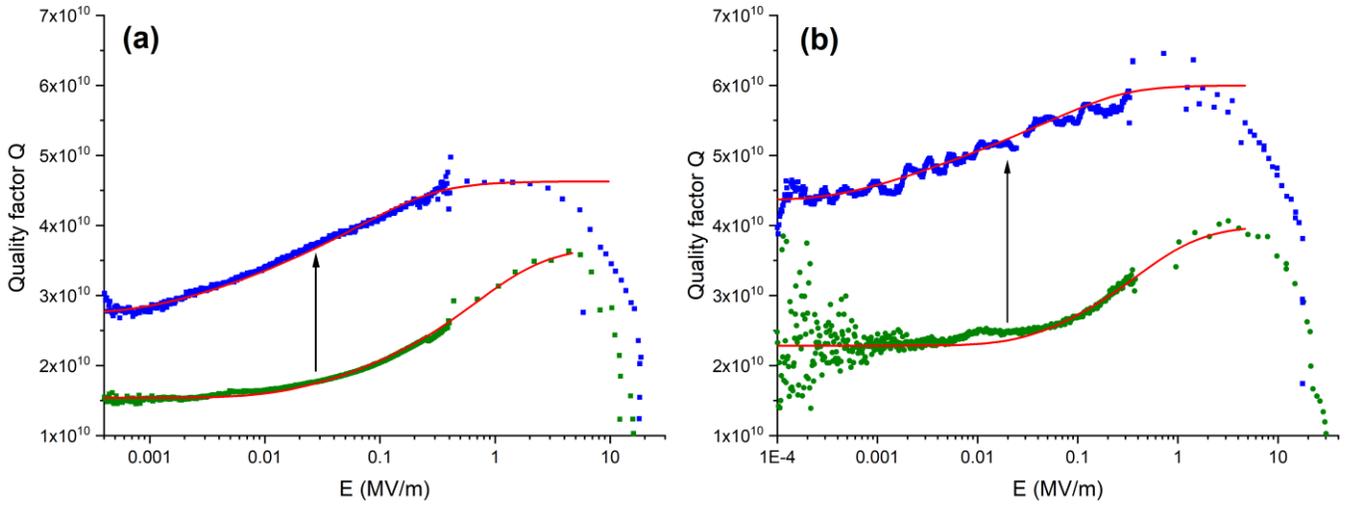

**Figure 1:** RF tests results showing the quality factor Q vs the accelerating gradient E at 1.5 K in the low field regime on a Nb cavity coated with $Al_2O_3$ (10 nm) and post-annealed at 650 °C for (a) 4 hours and (b) 10 hours. The baseline of EP cavity is in green and the curve of the $Al_2O_3$ coated cavities in blue. The red curves represent the TLS fit using equation 7 from [18].

The baseline RF tests (green curves) show typical performances after EP without post thermal treatments with Q values at low fields between 1.5 to $2.5 \times 10^{10}$. The blue curves represent the RF tests after deposition of 10 nm of $Al_2O_3$ and subsequent thermal treatments. It is clear that for both tests, the quality factor has been improved by a factor of two at low fields. The red lines are fits using the interacting and non-interacting two level system (TLS) model described in [18]. The fitting and the extracted TLS parameters are summarized in Table I.

**Table I :** Fitting parameters from equation 7 of ref. 18 and TLS parameters extracted from the fits.

| Treatments | Fitting parameters | | | | TLS parameters | | |
|---|---|---|---|---|---|---|---|
| | c ($C^2$/J) | ξ | $E_C$ (V/m) | $1/Q_{non-TLS}$ | $\sqrt{(T_1 T_2)}$ (s) | $\sigma_{TLS}$ ($cm^{-2}$) | $\tan(\delta_{TLS})$, ε, d(nm) |
| **Baseline EP** | $9.0\pm0.1\times10^{-24}$ | 140±10 | $2\pm1\times10^4$ | $2.7\pm0.1\times10^{-11}$ | $9.10^{-10}$ | $2.5\ 10^{11}$ | $1.5\ 10^{-3}$, 30, 5 |
| **$Al_2O_3$ 10 nm + 650 °C-4 hours** | $3.8\pm0.2\times10^{-24}$ | 500±50 | $9\pm1\times10^2$ | $2.1\pm0.1\times10^{-11}$ | $1.4\ 10^{-8}$ | $8.5\ 10^{10}$ | $7.7\ 10^{-4}$, 10, 10 |
| **Baseline EP** | $4.8\pm0.2\times10^{-24}$ | 40±2 | $3\pm1\times10^4$ | $2.5\pm0.1\times10^{-11}$ | $4.10^{-10}$ | $1.1\ 10^{11}$ | $6.9\ 10^{-4}$, 30, 5 |
| **$Al_2O_3$ 10 nm + 650 °C-10 hours** | $1.6\pm0.1\times10^{-24}$ | 750±50 | $4\pm1\times10^2$ | $1.7\pm0.1\times10^{-11}$ | $3.2\ 10^{-8}$ | $3.5\ 10^{10}$ | $3.2\ 10^{-4}$, 10, 10 |

The data and the fitting parameters reveal reproducible trends after $Al_2O_3$ deposition and annealing as compared to bare niobium with its native oxides:

1. The saturating electrical field $E_c$ is decreased by more than an order of magnitude. Using the relation $E_c = \sqrt{\frac{3}{2}} \frac{\hbar}{p} \frac{1}{\sqrt{T_1 T_2}}$ where $p \sim |e| \text{Å}$ is the electric dipole moment, and $T_1$, $T_2$ are the TLS energy relaxation and homogenous broadening times, we can deduce that $\sqrt{T_1 T_2}$ is increased by 15 to 80 times.

2. The spectral diffusion parameter, $\xi$ that describes the TLS impurities' coupling strength increases significantly.

3. The c parameter that describes the saturating value of the Q at very low electrical fields is also consistently reduced by a factor of 2 to 3 as compared to the baselines. Using the equations $c = \frac{\pi}{12} p^2 tanh\left(\frac{\hbar \Omega_0}{k_b T}\right) \rho'$ and $\tan(\delta_{TLS}) = \frac{4c}{\epsilon d} \frac{2 k_b T}{\hbar \Omega_0}$ in the limit $E, T \to 0$ and following the procedure of reference [18], the area density of TLS $\sigma_{TLS}$ and the loss tangent $\tan(\delta_{TLS})$ extracted from the fitting parameter c, the thickness of the oxide $d$ and the assumed dielectric constants $\epsilon$ listed in table I, are systematically reduced by a factor 2 as compared to the baselines.

These reproducible trends point toward a different nature of the TLS impurities in ALD coated and annealed Nb cavities as compared to native niobium oxides present in the cavity baselines. In order to investigate the microscopic origin of these changes in the RF performances and TLS losses, we performed STEM, HRTEM, EELS and EDX spectral imaging and XPS measurements on niobium samples that underwent the same processes as the cavities. Details on the apparatus, measurement conditions, analysis and fitting procedures can be found in the supplementary materials.

Fig. 2 summarizes the results obtained on a cavity-grade electro-polished Nb sample coated with 100 cycles of $Al_2O_3$ by ALD without any thermal post-treatment. The high-resolution STEM images show an amorphous 12 nm thick $Al_2O_3$ film (light grey) on top of an intermediate 5 nm amorphous layer (darker grey hue) on crystalline Nb (Fig. 2(a)) [15,19]. The EELS analysis (Fig. 2(b)) reveals that this 5 nm interfacial layer is composed of $NbO_x$ (See the SI references for details on how the $NbO_x$ component is isolated). The XPS allows for a more detailed chemical analysis and the Nb-3d core levels spectrum shows that the $NbO_x$ is composed mostly of $NbO_2$ (7 %), $NbO$ (24 %) and $Nb_2O_5$ (23 %) (in Fig. 2(c) bottom). This Nb oxide composition differs significantly from a typical un-coated electro-polished Nb sample with a native oxide composition dominated by $Nb_2O_5$ (57 %) and displayed in Fig. 2(b) top for comparison. The partial reduction of the $Nb_2O_5$ to sub-oxides is caused by the ALD deposition temperature of 250 °C applied for 2-3 hours in agreement with previous works [20].

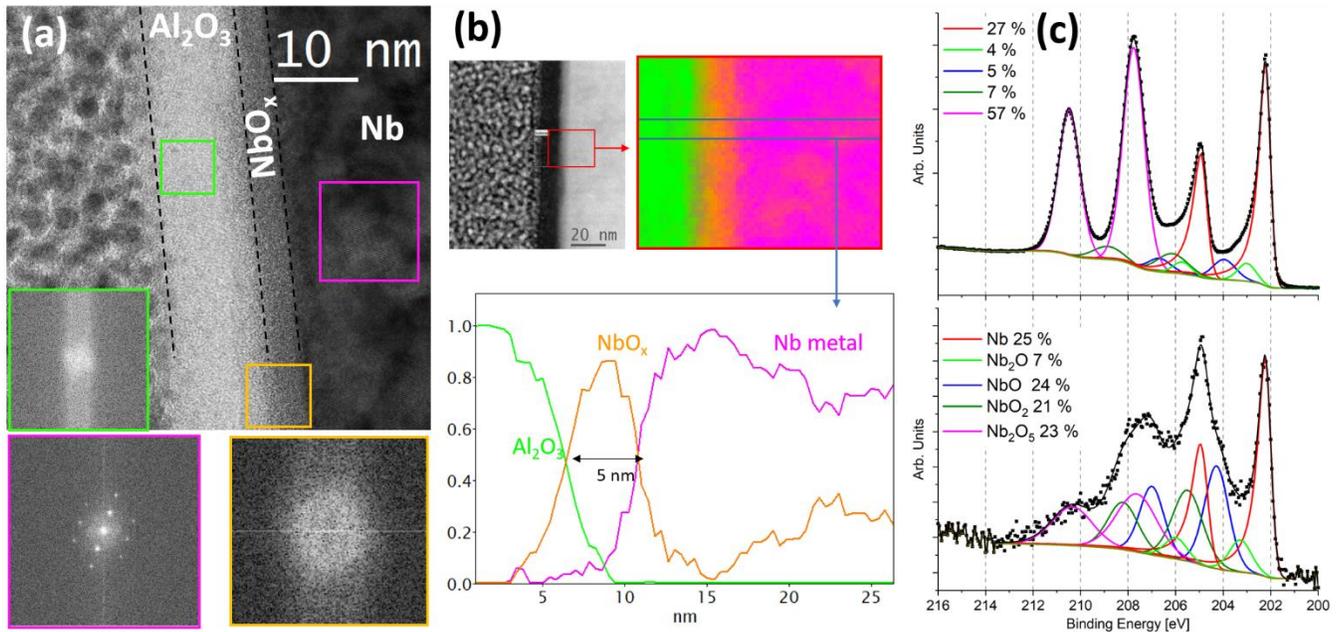

**Figure 2:** a) Bright field STEM imaging and local FFT analyses, b) HAADF(top left) and EELS analysis of an as-deposited Al$_2$O$_3$-coated Nb and (c) XPS spectra of Nb-3d core levels of a reference EP niobium sample (top). XPS spectra of Nb 3d core level of as deposited Al$_2$O$_3$-coated Nb (bottom).

After annealing the Al$_2$O$_3$-coated Nb sample at 650 °C for 4 hours, the high resolution STEM and EELS measurements (Fig. 3 (a)) reveal an amorphous 10 nm Al$_2$O$_3$ layer on top of an amorphous 1.5-2 nm thick Nb oxide interface. The composition of this oxide measured by XPS (Fig. 3(b)) is NbO (14 %) and Nb$_2$O (5 %). The thickness reduction of the Nb oxide layer from 5 to 1.5-2 nm along with a strong decrease in total Nb oxide composition from 75 % to 19 % of the Nb-3d spectrum after annealing indicates a further reduction and a diffusion of the oxygen from the Nb oxide layer into the bulk Nb.

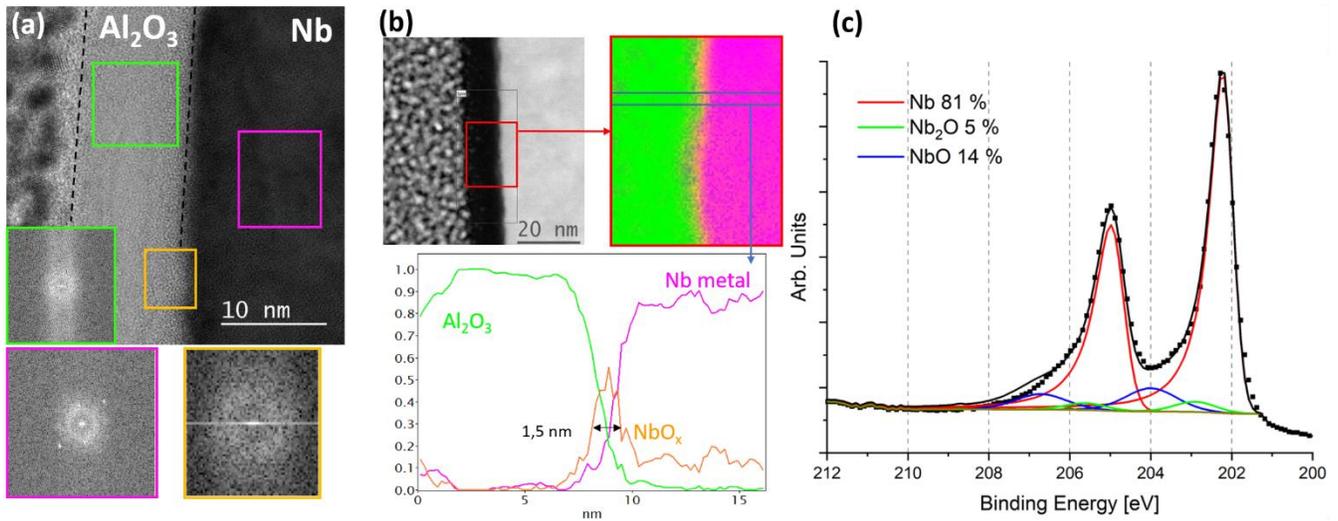

**Figure 3: a) Bright field STEM imaging and local FFT images b) HAADF (top left) and EELS analysis and c) XPS spectrum of Nb-3d core levels of an Al$_2$O$_3$-coated Nb sample after annealing at 650 °C during 4 hours.**

Upon increasing the annealing time from 4 to 10 hours at 650 °C, the TEM and EDX analyses (Fig. 4(a)) show that the Al$_2$O$_3$ thickness remains unchanged at 10 nm. The Nb oxide layer however, becomes discontinuous, keeping a thickness of about 2 nm in regions where it is still present. The XPS data analysis reveals an increased Nb$_2$O concentration from 5 % to 37 % whereas the NbO decreases from 14 % to 7 %. The decrease of NbO and increase of Nb$_2$O concentrations indicate a further reduction of NbO into Nb$_2$O upon longer annealing time but it cannot account however for the total concentration of Nb$_2$O measured. We suspect that the Al$_2$O$_3$ interface with the Nb could start releasing partially its oxygen atoms to the Nb underneath under the combined influence of prolonged annealing and high Nb reactivity with O.

The HRTEM images display the crystalline structure of bcc Nb with a strong structural elongation (2-3 %) perpendicular to the surface and indicate local areas with larger d-spacing (0.238-0.241 nm) in comparison to the bulk Nb (0.229-0.236 nm). This distortion could be associated with a local increase in the crystal parameter, related to the inclusion of oxygen into the niobium unit cell, as supported by the XPS data and seen previously [21].

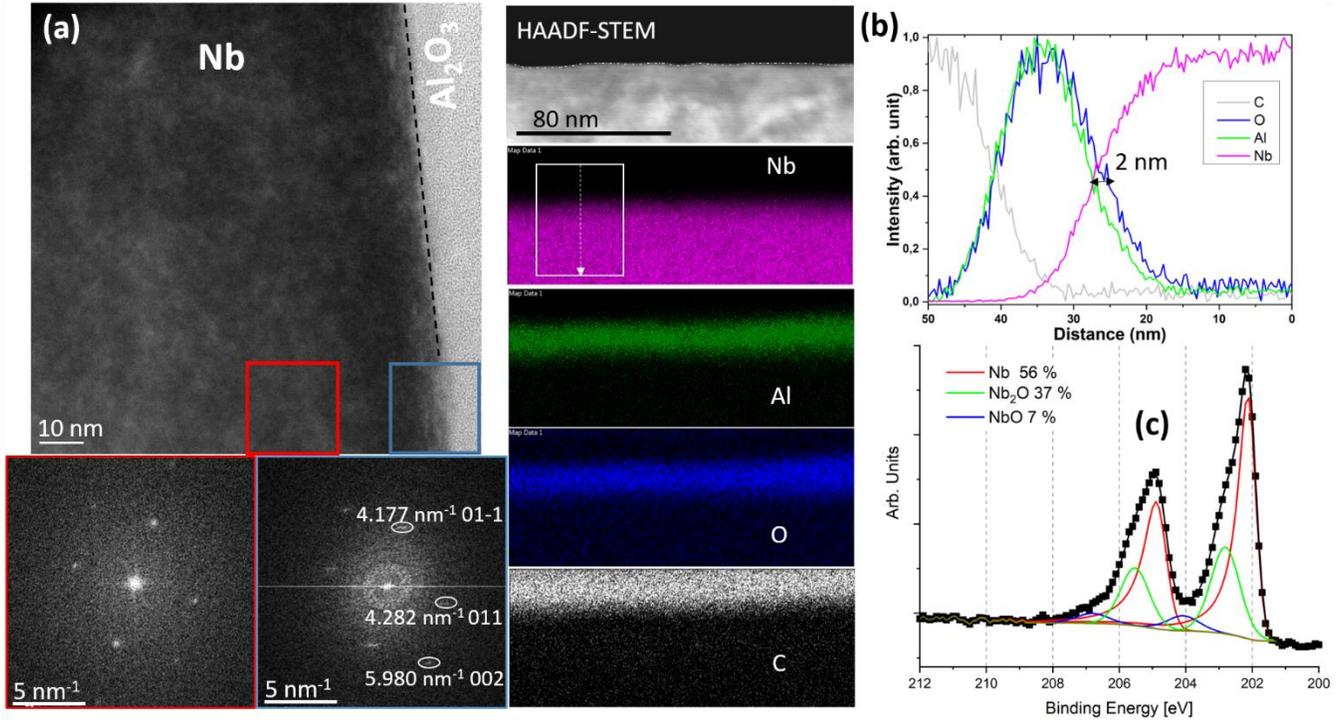

**Figure 4:** a) HRTEM and local FFT images showing the distances and the corresponding lattice planes in the reciprocal space, b) EDX analyses and c) XPS spectrum of Nb-3d core levels of an $Al_2O_3$-coated Nb sample after annealing at 650 °C during 10 hours.

The baseline RF performances at low field are typical for EP Nb cavities, it is therefore reasonable to assume that the Nb oxide thickness and structure are similar to what was found in [18], i.e, a 5 nm thick amorphous $Nb_2O_5$ layer on top of a ~ 1.5-2 nm sub-oxide $NbO_x$ layer. Our microscopic analyses of the $Al_2O_3$-coated and annealed niobium surface show a similar $NbO_x$ thickness and composition whereas a 10 nm amorphous $Al_2O_3$ film replaces the amorphous $Nb_2O_5$. The $Nb_2O_5$ removal can therefore explain the reproducible improvement of the quality factor at low RF field amplitudes upon annealing at 650 °C for 4 hours or 10 hours and the corresponding reduction by a factor of 2 of the TLS losses $\tan(\delta_{TLS})$ listed in Table I. This result is in agreement with previous works [4, 6] that emphasize the strong contribution and dependence of RF TLS losses on the thickness of $Nb_2O_5$ in SRF cavities. Furthermore, the predominant presence of the amorphous ALD $Al_2O_3$ layer also explains the notable changes in the nature of the previously mentioned TLS defects that entail very different values as compared to $Nb_2O_5$ for the critical saturation field $E_c$ and the measured coupling strength $\xi$.

Previous measurements of the loss tangent of ALD-deposited $Al_2O_3$ films on superconducting resonators [22] give values of ~ 2-3×$10^{-3}$ for films ranging from 30 to 100 nm whereas our values are about one order of magnitude lower: ~ 3×$10^{-4}$. The HRTEM analysis reveals that the $Al_2O_3$ film thickness changes from 12 to 10 nm upon annealing, indicating a densification (in the absence of measurable Al diffusion into the Nb) from $3\pm0.1$ g/cm$^3$ as measured by X-ray reflectivity (XRR) on as-deposited films on Si samples (supplementary information), to an estimated 3.5 g/cm$^3$ on niobium after the thermal treatments. In-depth investigation of ALD synthesized $Al_2O_3$ films showed that a significant concentration of hydroxyl groups remain in the film during the growth, contributing to the low film density [23], that are fully eliminated by post-annealing above 1000 °C [24]. The hydroxyl groups have been proposed, and in some cases identified [25, 26, 27, 28], as a potential source of TLS losses. Their concentration decrease during the annealing step in HV at 650 °C could provide a microscopic origin for the lower loss tangent in our annealed films as compared to the as-deposited $Al_2O_3$ measured in [22].

In order to reduce further the niobium sub-oxide presence at the interface, we have increased the post-annealing temperature to 800 °C for few hours. Subsequent depth-profile XPS measurements revealed the presence of Nb and O (at the surface) without Al, indicating that the $Al_2O_3$ film was no longer present at the surface and that it did not diffuse deeper into the Nb (not shown). We suspect that the oxygen diffusion mechanism proposed earlier that emerges at 650 °C during 10 hours becomes more severe at 800 °C and the progressive reduction of the $Al_2O_3$ film to metallic aluminum causes its evaporation into the vacuum chamber due to its low melting temperature of 660 °C.

In conclusion, we have investigated the effect of $Al_2O_3$ deposition by ALD and post annealing on the RF performances at low fields of Nb 1.3 GHz resonators. This approach resulted in an enhancement of the low-field quality factor caused by a modification of the nature of TLS defects at the surface and a reduction of the loss tangent associated with important modifications of the niobium native oxide chemical composition and structure, as measured by STEM-EELS, HRTEM and XPS. The creation of a protective $Al_2O_3$ layer at the surface enables air-stable performance improvements and facilitates greatly 3D superconducting resonator handling and characterizations. The combined deposition and thermal treatment approach provides a platform to study the effects of passivation layers with different chemical and structural (crystalline, amorphous, thickness) properties on the TLS losses and on the nature of the interface superconductor/dielectric defects.

**SUPPLEMENTARY MATERIAL**

**X-ray reflectivity measurement (XRR)**

XRR measurements where performed using a 5-circle diffractometer (Rigaku SmartLab) with CuKα radiation from a rotating anode. The curves were fitted using the "X'pert reflectivity" software to determine the thickness, density and roughness of deposited films. Fig. 5 shows the fit using a density of 3.1 g/cm$^3$, an $Al_2O_3$ thickness of 9.5 nm and a roughness of 0.8 nm.

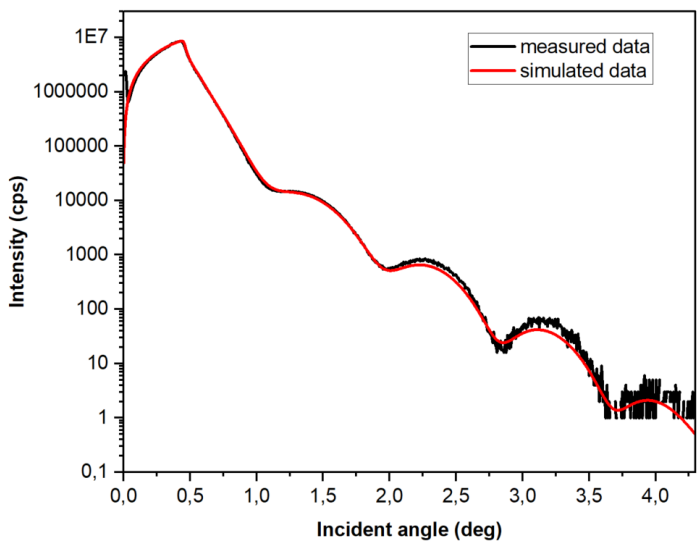

**Figure 5: XRR analysis of the $Al_2O_3$ film deposited on Si**

**X-ray photoelectron spectroscopy measurement and analysis (XPS)**

XPS measurements presented in Fig. 2, Fig. 3 and Fig. 4 were acquired using K Alpha spectrometers equipped with a monochromatic Al $K_\alpha$ excitation (1486.6 eV) and charge compensation systems. The binding energies were calibrated against the C1s binding energy set at 284.8 eV. The spectra were treated by means of CasaXPS software [29] after the subtraction of a Shirley-type background. In order to have a better resolution of the Nb 3d level spectra, Ar+ ions was used to partially etch the $Al_2O_3$ film. The lineshapes used to fit the symmetrical peaks were mixed Gaussian-lorentzian, GL(30), and an asymmetrical lineshape, A(0.38,0.6,10)GL(10), for the metallic component of Nb3d core-level spectrum.

**Table II: XPS fitting parameters of the 3d Nb core-level spectrum [30]**

| Component | Peak position (eV) | FWHM (eV) |
|---|---|---|
| Nb | 202.2 (202, 202.5) | 0.55 (0.4, 0.7) |
| $Nb_2O$ | 203.1 (202.8, 203.4) | 1 (0.75, 1.2) |
| NbO | 204.3 (203.4, 204.7) | 1.1 (0.75, 1.2) |
| $NbO_2$ | 205.5 (205.2, 206.5) | 1.5 (0.75, 1.5) |
| $Nb_2O_5$ | 207.6 (207.5, 207.8) | 1.8 (1, 2) |

**Transmission electron microscope analysis (TEM)**

In order to investigate by TEM the nature of the interface between the Nb and the protective alumina layer, cross section lamellae were prepared in a FIB microscope and preliminarily studied by TEM at JaNNus-Orsay,IJCLab. Note that for the FIB preparation, the sample surface was firstly protected by a deposition of carbon followed by another layer of platinum and a standard lift-out procedure was performed using a Ga ion source.

In Fig. 2 and Fig. 3, bright field and high-angle annular dark-field images and electron energy-loss spectroscopy (EELS) spectrum-images of the cross-section samples were acquired at LPS using a Nion Ultrastem 200 scanning transmission electron microscope (STEM) operating at 100 kV. The probe convergence angle was 35mrad. The data shown here used a 50mrad EELS collection angle. EELS spectrum-images were processed by a spectral unmixing method. Within the spectral unmixing framework, individual EELS spectra, each corresponding to a unique pixel, is represented as a linear combination of characteristic spectral signatures along with their respective fractional weights [31]. The application of unmixing enables the exploitation of information embedded in the fine structure of the edges and to achieve more advanced results than can be obtained by elemental analysis alone. In this study it was thus possible to discern oxygen contributions linked to $NbO_x$ from those associated with $Al_2O_3$. Notably, the O-K edge's fine structure differs between these two compounds. The unmixing procedure employed the Vertex Component Analysis (VCA) method [32], which has demonstrated efficacy in the unmixing of STEM-EELS data.

In Fig. 4, TEM characterizations on a lamella were conducted using a ThermoFisher Titan-Themis 300 G3 operating at 300 kV. This microscope is equipped with a BM-Ceta- camera for HRTEM imaging, high-angle annular dark-field (HAADF) and bright field (BF) detectors for scanning TEM (STEM) imaging and a silicon drift detector (SDD) from Oxford instrument for Energy Dispersive X-ray (EDX) analyses. To study the elemental distribution, HAADF-STEM images were combined with EDX analyses, enabling to identify the presence of Al, O and Nb in the sample. Note that Cu was detected and originates from the TEM grid. Further analyses of on EDX data were realized by normalization of the line profile extracted from the individual elemental map.

## ACKNOWLEDGMENTS

The authors would like to thank Enrico Cenni from Commissariat de l'Energie Atomique (CEA) for providing the electrical field distribution in a 1.3 GHz cavity for the TLS numerical simulations, Mohammed Fouaidy and Thierry Pepin Donat from IJCLab for providing a HV thermal treatment and Claire Antoine from CEA for the insightful discussions. This work was partly supported by the French RENATECH network (Focused Ion Beam TEM thin foils preparation by David Troadec at IEMN Lille) and by the CNRS-CEA METSA French network (FR CNRS 3507) for the STEM experiments at LPS). This project has received funding from the region Ile de France project SESAME AXESRF, the European Union's Horizon 2020 Research and Innovation program under Grant agreement No 101004730 and Grant agreement No 730871.

## AUTHORS DECLARATIONS

**Conflict of Interest:** Authors Y.K and T.P. have patent N° PCT/FR2023/051937 pending.

**Authors Contributions: Yasmine Kalboussi:** Conceptualization (equal); Data curation (equal); Formal analysis (equal); Investigation (equal); Methodology (equal); Resources (equal); Validation (equal); Visualization (equal); Writing–original draft (equal). **Baptiste Delatte**: Resources (equal). **Sarra Bira**: Resources (equal). **Kassiogé Dembele**: Investigation (equal); Formal analysis (equal); Data Curation (equal); Writing-Reviewing and Editing (equal). **Xiaoyan Li**: Investigation (equal); Formal analysis (equal); Data Curation (equal). **Frederic Miserque**: Investigation (equal); Writing-Reviewing and Editing (equal). **Nathalie Brun** : Investigation (equal); Formal analysis (equal); Data Curation (equal); Writing-Reviewing and Editing (equal). **Michael Walls**: Investigation (equal); Formal analysis (equal); Data Curation (equal); Writing-Reviewing and Editing (equal). **Jean-Luc Maurice**: Investigation (equal); Formal analysis (equal); Data Curation (equal). **Diana Dragoe:** Investigation (equal). **Jocelyne Leroy**: Investigation (equal)**. David Longuevergne:** Resources (supporting). **Aurélie Gentils**: Investigation (equal). **Stéphanie Jublot-Leclerc**: Investigation (equal). **Gregoire Julien:** Resources (equal)**. Fabien Eozenou:** Resources (equal)**. Matthieu Baudrier:** Investigation (equal)**. Luc Maurice:** Investigation (equal)**. Thomas

**Proslier:** Conceptualization (equal); Data curation (equal); Formal analysis (equal); Funding Acquisition (lead); Supervision (lead); Investigation(equal); Methodology(equal); Validation (equal); Visualization (equal); Writing–original draft (equal).

## DATA AVAILABILITY

The data generated and analyzed here are available from the corresponding authors upon reasonable request.